\documentclass[onecolumn]{aastex62}

\usepackage{xcolor}
\received{July, 2018}
\revised{}
\accepted{}
\shorttitle{Photo-desorption of circumstellar ice}
\shortauthors{Jim\'enez-Escobar et al.}

\begin{document}

\title{Photo-desorption of H$_2$O:CO:NH$_3$ circumstellar ice analogs: Gas-phase enrichment}

\author{A. Jim\'enez-Escobar, A. Ciaravella, C. Cecchi-Pestellini}
\affil{INAF - Osservatorio Astronomico di Palermo, P.za Parlamento 1, 90134 Palermo, Italy}
\author{C.-H. Huang, N.-E. Sie, Y.-J. Chen}
\affil{Department of Physics, National Central University, Jhongli City, Taoyuan County 32054, Taiwan}
\author{G.M. Mu\~{n}oz Caro}
\affil{Centro de Astrobiolog\'ia (INTA-CSIC), Carretera de Ajalvir, km 4, Torrej\'on de Ardoz, 28850 Madrid, Spain \\} 

\begin{abstract}
We study the photo-desorption occurring in  H$_2$O:CO:NH$_3$ ice mixtures irradiated with monochromatic (550 and 900 eV) and  broad band (250--1250 eV) soft X-rays generated at the  National Synchrotron Radiation Research Center (Hsinchu, Taiwan). We detect many masses photo-desorbing, from atomic hydrogen (m/z = 1) to complex species with m/z = 69 (e.g., C$_3$H$_3$NO, C$_4$H$_5$O, C$_4$H$_7$N), supporting the enrichment of the gas phase. 

At low number of absorbed photons, substrate-mediated exciton-promoted desorption dominates the photo-desorption yield inducing the release of weakly bound (to the surface of the ice) species; as the number of weakly bound species declines, the photo-desorption yield decrease about one order of magnitude, until porosity effects, reducing the surface/volume ratio, produce a further drop of the yield.  

We derive an upper limit to the CO photo-desorption yield, that in our experiments varies from 1.4 to 0.007 molecule photon$^{-1}$ in the range $\sim 10^{15} - 10^{20}$~absorbed photons cm$^{-2}$. We apply these findings to a protoplanetary disk model irradiated by a central T~Tauri star. 
\end{abstract}

\section{Introduction}\label{intro}
Interstellar ice mantles, formed mainly by H$_2$O, may also contain a variety of C$-$ and N$-$bearing molecules such as CO$_2$, CO, CH$_3$OH, and NH$_3$ \citep{B15}. Laboratory simulations indicate that energetic processing of these ices can produce a large amount of complex organic compounds, which can include molecules of biological interest such as alcohols, quinones, esters, amino acids (e.g., \citealt{Mu02}), and sugars \citep{Me16}. Recently,  during the Rosetta mission  many complex organic molecules such as ethanol (CH$_3$CHO) and formamide (NH$_2$CHO) have been detected in the  environment of comet 67P/Churyumov-Gerasimenko  \citep{Goe15}. In the accompanying coma, mass spectroscopy measurements are compatible with biologically active compounds, such as glycine \citep{Alt16},  or glycine structural isomers \citep{O16}.

Gas phase observations  have  detected a large number of molecular species in cold clouds in which the temperature is around $\rm 10~K$. Thus, non thermal desorption from dust surface are crucial for  our understanding of the chemical evolution of  such regions. In addition to photon-induced desorption, cosmic-ray bombardment of icy dust is expected to drive non-thermal desorption. Because of the low penetration depth of ultraviolet photons, only the outskirts of dense clouds are fully processed by ultraviolet; well inside the cloud, ice mantles are processed by the secondary ultraviolet field generated by cosmic-ray excitation of molecular hydrogen. The direct impact of cosmic rays on icy dust is expected to desorb volatiles significantly \citep{D15}, while X-rays may also contribute to the desorption of ice mantles in circumstellar regions aroung young stars. Release of chemical energy was also proposed as a non-thermal desorption mechanism (e.g., \citealt{G07}). Photochemical desorption or photo-chemidesorption is the inmediate desorption of molecules triggered by a photon after their formation on the ice surface, which leads to a rather constant desorption rate \citep{Mar15}. Ice explosions could be driven by chemical energy, released sporadically when the growing population of radicals in the ices spontaneously associated (e.g., \citealt{R13}).  

Ultraviolet radiation can induce desorption from the surface through two principal mechanisms. The first one, Desorption Induced by Electronic Transitions (DIET; e.g., \citealt{Mar15}), is initiated by absorption in the first layers of the ice, exciting molecules to high-laying electronic states. The excitation energy is then dissipated to the surface of the ice by transferring to neighbouring molecules. The transmitted energy is high enough to overcome  intermolecular interactions, eventually ejecting the molecule into the gas-phase \citep{Ob07, Mu10, Fay11, Fay13, Mar15}. The second mechanism is photo-chemi-desorption \citep{Mar15,Mar16}.

X-rays are absorbed deep inside the ice matrix, leading to the fragmentation of molecules. In general, this occurs through the ionization or excitation of an inner-shell electron, followed by either normal or resonant Auger decays, with the normal Auger decay active above the ionization threshold. The injection of energetic photoelectrons produces multiple ionization events generating a secondary electron cascade that dominates the chemistry (e.g., \citealt{Jim16}).  X-ray induced photo-desorption may be caused by three channels: (1) Photon Stimulated Desorption (PSD) resulting from the fragmentation of surface molecules after their own core excitation \citep{Par02}; (2) Auger-Stimulated Ion Desorption (ASID) where after core electron excitation, Auger transitions cause the formation of two holes in the valence orbital, with final desorption of the resulting ion after relaxation \citep{Sek97}; (3) X-ray induced Electron-Stimulated Desorption (XESD) in which electrons emitted from the bulk can induce non-resonant fragmentation of the surface molecules by, e.g., interband transition, and dissociative electron attachment. In that case, the fragmentation is simply proportional to the electron yield \citep{Par02}. Non-thermal desorption of molecular ices may be mediated by excitons, in terms of excitation of a substrate-adsorbate complex and the Menzel-Gomer-Redhead mechanism \citep{MC18}. 

In this work we study the photo-desorption occurring in  H$_2$O:CO:NH$_3$ ice mixtures irradiated with monochromatic (550 and 900 eV) and  broad band (250--1250 eV) soft X-rays from the BL08B beamline at the  National Synchrotron Radiation Research Center (NSRRC, Hsinchu, Taiwan). Other effects of soft X-ray irradiation, i.e. formation of volatile photo-products and the organic residue that remains at room temperature, are discussed in a companion paper \citep{Cia18}.

The experiments are described in Section~\ref{protocol}, photo-desorbing species induced by X-rays are listed and discussed in Section~\ref{photo}, and the dependence of photo-desorption yield on the physical conditions during the experiment (e.g., the photon rate) in Section~\ref{anal}. Results are discussed in Section \ref{ph_2}. Conclusions and astrophysical implications of X-ray induced photo-desorption are presented in the last Section.

\section[Experiments]{Experiments} \label{protocol}
The experiments were carried out in the Interstellar Photo-process System (IPS), an ultrahigh vacuum (UHV) chamber of base pressure $1.3\times10^{-10}$ mbar. A  Fourier Transform Infrared ABB FTLA-2000-104 spectrometer equipped with a mercury-cadmium-telluride infrared detector  and a Quadrupole Mass Spectrometer (QMS) in the range of $1 - 200$ amu range (0.5 amu resolution) were used to monitor the ice and the composition of the gas in the chamber. The gas line system baked out at $120~\rm ^\circ C$  to eliminate organic and water contamination reaches a minimum pressure of $1.3 \times 10^{-7}$ mbar before preparing the gas mixture for the experiments. For a detailed description of the IPS facility  we refer to the work of \citet{Ch14}. 

We irradiated an ice mixture H$_2$O+CO+NH$_3$ (1.5:1.3:1 in ratio) at 15 K for a total of 120 min. We choose a mixture rich in CO and NH$_3$ to maximize the photo-products abundances otherwise difficult to be detected. As X-ray source we used the spherical grating monochromator beamline BL08B at NSRRC covering photon energies from 250 to 1250 eV. The spectrum is the same as in \citet{Cia18}. We irradiated with the whole available band (Broad Band, BB) at two different fluxes, and monochromatic photons of 550 (M5), and 900 (M9)~eV. Before and after irradiation, infrared spectra were collected with a resolution of 1, 2, and 4 cm$^{-1}$. During the experiment, the irradiation was stopped several times ($\Delta t = 0.5,1,5,10,20,40,60,80,100$, and 120~min), when infrared spectra were taken with a resolution of 2 cm$^{-1}$. At the end of the irradiation, the ice was heated  up to room temperature at a rate of 2~K~min$^{-1}$. During the warm-up, infrared spectra were acquired every 10~K with a resolution of 4 cm$^{-1}$. 

The set of experiments are reported in Table 1, in which for each experiment we also report the employed X-ray energies, initial H$_2$O, CO and NH$_3$ column densities, photon rate, the total number of impinging and absorbed photons in the ice, and the total absorbed energy by the sample in each experiment. The X-ray flux impinging on the ice decreases exponentially as it propagates inside the ice. To compute such decrease we exploit the X-ray database tool (http://henke.lbl.gov) that includes photoabsorption, scattering, and transmission coefficients by \citet{Hen93}. The photon flux has been derived taking into account the size of the X-ray spot, $A_{\rm X} = 0.08$~cm$^2$.
\begin{deluxetable*}{ccccccccc}
\label{tone}
\centering
\tablecaption{Irradiation Experiments}
\tablewidth{0pt}
\tablehead{
Exp. & Energy & N(H$_2$O) & N(CO) & N(NH$_3$) & Photon rate & Tot. imp. ph. & Tot. abs. ph.& Tot. abs. E.\\
        &  (keV)  & (cm$^{-2}$) & (cm$^{-2}$) & (cm$^{-2}$) & (photon s$^{-1}$) & (photons) & (photons) & (eV)}
\startdata
M5 & 0.55           & 4.34$\times$10$^{17}$ & 3.57$\times$10$^{17}$ & 2.95$\times$10$^{17}$ & 9.02$\times$10$^{12}$ & 6.49$\times$10$^{16}$ & 3.02$\times$10$^{16}$ & 1.66$\times$10$^{19}$ \\
M9 & 0.90           & 4.52$\times$10$^{17}$ & 3.56$\times$10$^{17}$ & 2.70$\times$10$^{17}$ & 7.75$\times$10$^{12}$ & 5.58$\times$10$^{16}$ & 9.07$\times$10$^{15}$ & 8.13$\times$10$^{18}$ \\
BBw & 0.25--1.2 & 4.42$\times$10$^{17}$ & 3.57$\times$10$^{17}$ & 2.79$\times$10$^{17}$ & 6.62$\times$10$^{13}$ & 4.77$\times$10$^{17}$ & 1.14$\times$10$^{17}$ & 7.56$\times$10$^{19}$ \\
BBs & 0.25--1.2  & 4.15$\times$10$^{17}$ & 3.54$\times$10$^{17}$ & 2.72$\times$10$^{17}$ & 6.03$\times$10$^{15}$ & 4.34$\times$10$^{19}$ & 1.01$\times$10$^{19}$ & 6.67$\times$10$^{21}$ 
\enddata
\end{deluxetable*} 

The column density values reported in Table~1 are obtained using the following band strengths: $A \rm (H_2O) = 1.2 \times 10^{-17}$ cm molecule$^{-1}$ \citep{Ger95} at 1660 cm$^{-1}$, $A \rm (CO) = 1.1 \times 10^{-17}$~cm molecule$^{-1}$ \citep{Jia75} at 2142 cm$^{-1}$, and $A \rm (NH_3) = 1.7 \times 10^{-17}$ cm molecule$^{-1}$ \citep{San93} at 1112 cm$^{-1}$. The total column density of the mixture is computed as the sum of the column densities of the components, $N_{\rm T} = N_{\rm H_2O} + N_{\rm CO} + N_{\rm NH_3}$.

We plot in Figure~1 the infrared spectra of the BBs ice mixture before and after irradiation of 120 min equivalent to a total impinging photons of  $4.34 \times 10^{19}$ (see Table~1). The irradiation with different photon rates leads to the same photo-products. In this paper, only the photodesorbed species are reported. A complete list of the products is in Table 1 of \citet{Cia18}.  
\begin{figure}
\label{fone}
\centering
\includegraphics[width=18cm]{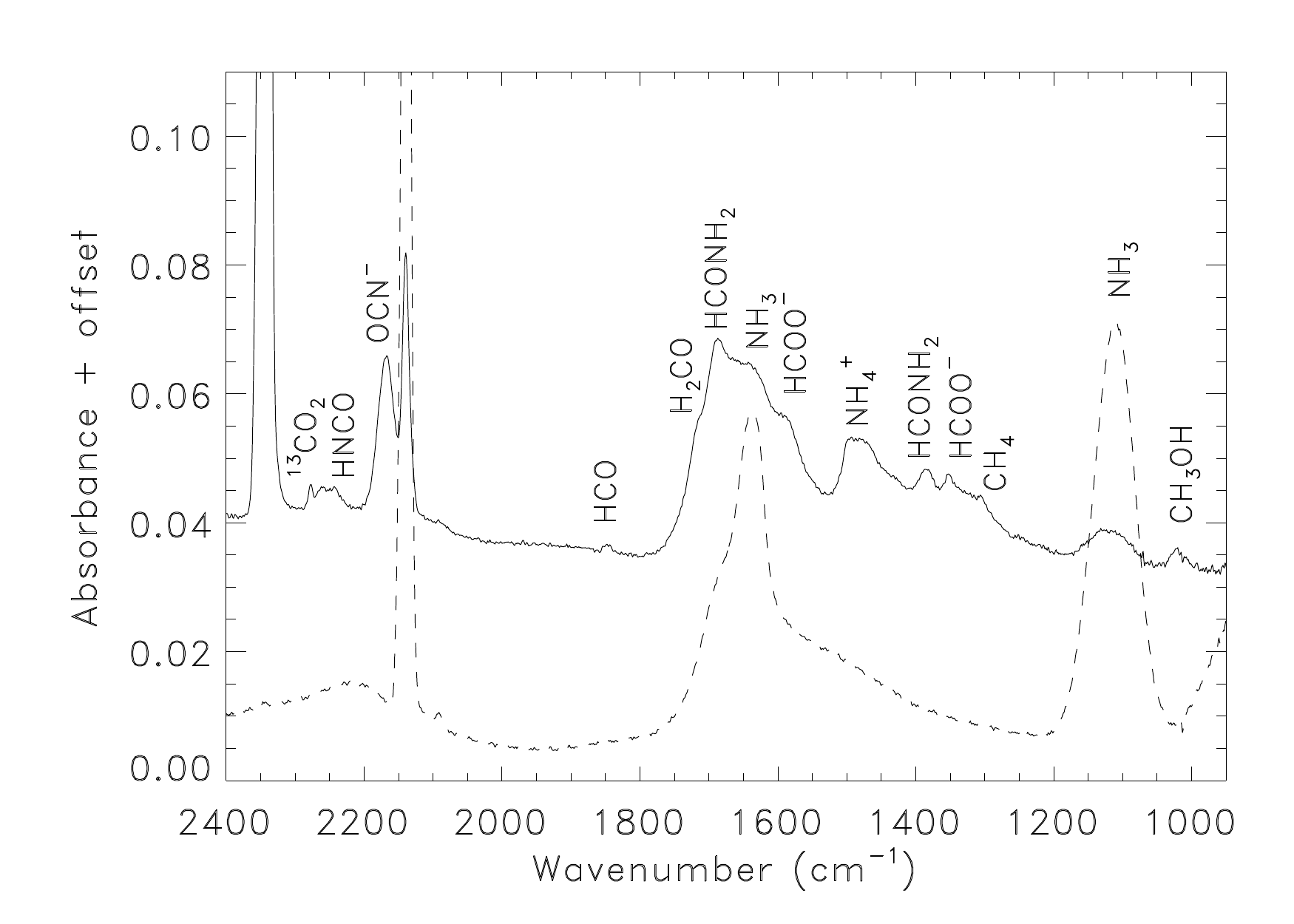} 
\caption{Infrared spectra of the BBs ice sample before (dashed line) and after (solid line) irradiation. The irradiated spectrum has been shifted for clarity purpose.}
\end{figure}

\section{Photo-desorbing fragments induced by X-rays} \label{photo}
In this work we study the photo-desorption arising during the X-ray irradiation of the ice sample, exploring different radiation energies and fluxes. The assumed fluxes cover about three orders of magnitude. 

X-rays induce photo-desorption of the ice mixture components (parent molecules) as well as of the products. Irradiation with the weakest fluxes (M5 and M9) produced only a faint desorption of mass 28. In the BBw experiment, desorptions of masses 1, 2, 12, 17, 18, 28, 30, 32, and 44 have been observed, while the more intense photo-desorption and the largest mass inventory have been detected using the strongest flux (BBs). For such reason, we mainly focus on the BBs experiment. Figure~2 shows all the masses photodesorbing during the BBs experiment. 

\begin{deluxetable}{c c c c }
\label{ttwo}
\tabletypesize{\scriptsize} 
\tablecaption{ Fragmentation of the parent molecules}
\tablewidth{0pt}
\tablehead{ Fragments&  \multicolumn{3}{c}{Species}\\
 (m/z)  &  CO &  H$_2$O & NH$_3$}
\startdata 
12 & 0.04414 &              &                 \\        
14 &               &              &    0.022    \\
15 &               &              &    0.071    \\
16 & 0.01458 &  0.021   & 0.8948     \\  
17 &               & 0.259    &  1             \\
18 &               &    1        &                 \\
19 &               &  0.0011 &                 \\
28 & 1            &              &                 \\
\enddata
\end{deluxetable} 
All masses show 10 desorption peaks, as many as the number of radiation steps. The intensity of the desorption peaks decreases rapidly with the irradiation time. In Table~2 we list 
the fragmentation pattern of parent molecules produced by electron impact ionization in the QMS. In Figure~3 we plot for each mass the total ion current after each radiation step against the irradiation time. The total ion current is computed as sum of all accumulated desorption peaks up to the selected radiation step. The signal of QMS spectrum extends over more than three orders of magnitude, and is dominated by m/z = 2, 18 and 28 over any other species.  
\begin{figure}
\centering
\includegraphics[width=18cm]{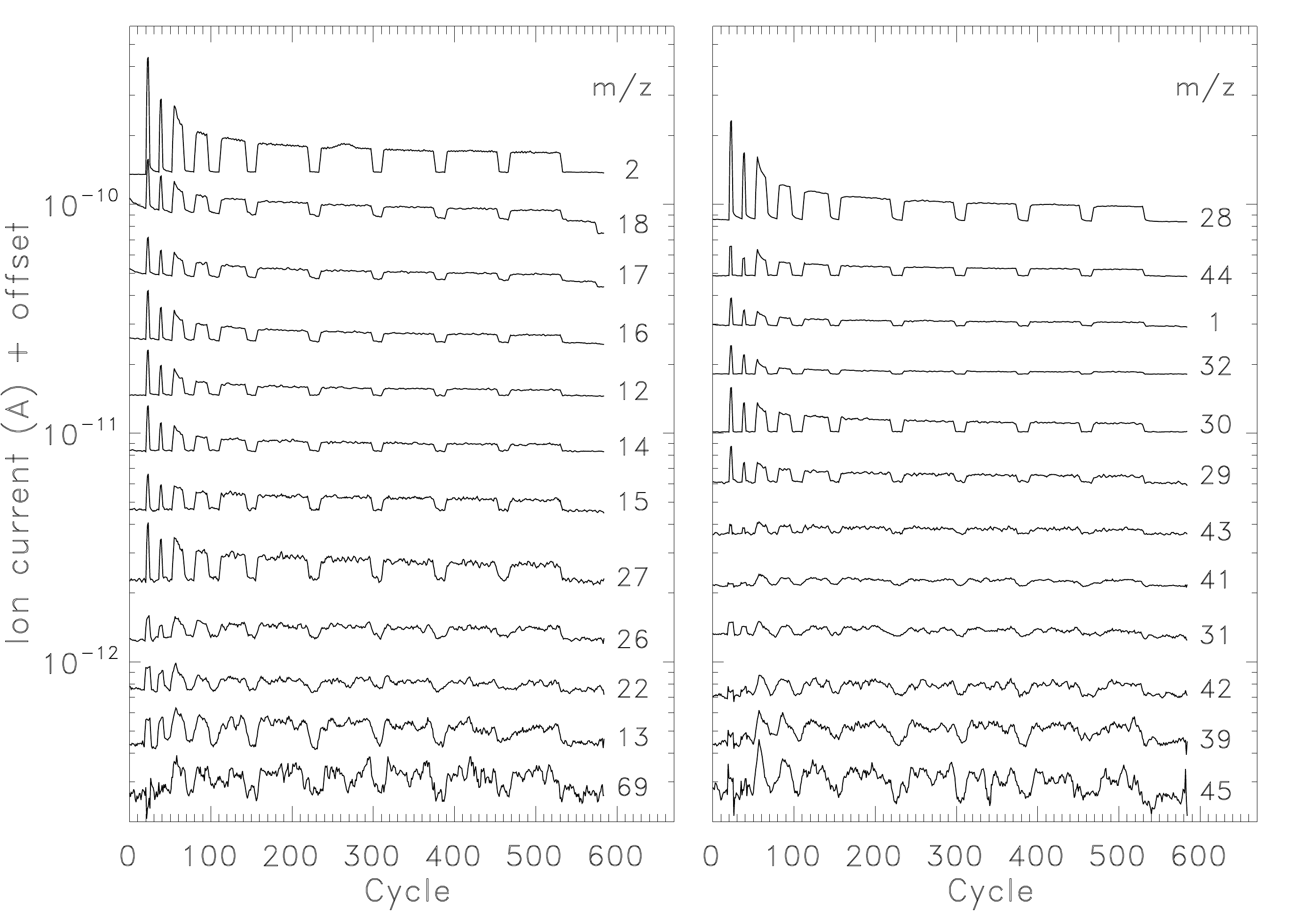} 
\caption{Ion current for the masses detected by QMS during BBs experiment as a function of the scanning time measured in cycles (one cycle is about 19~s). The curves have been shifted for clarity purpose.}
\label{ftwo}
\end{figure}
\begin{figure}
\centering
\includegraphics[width=18cm]{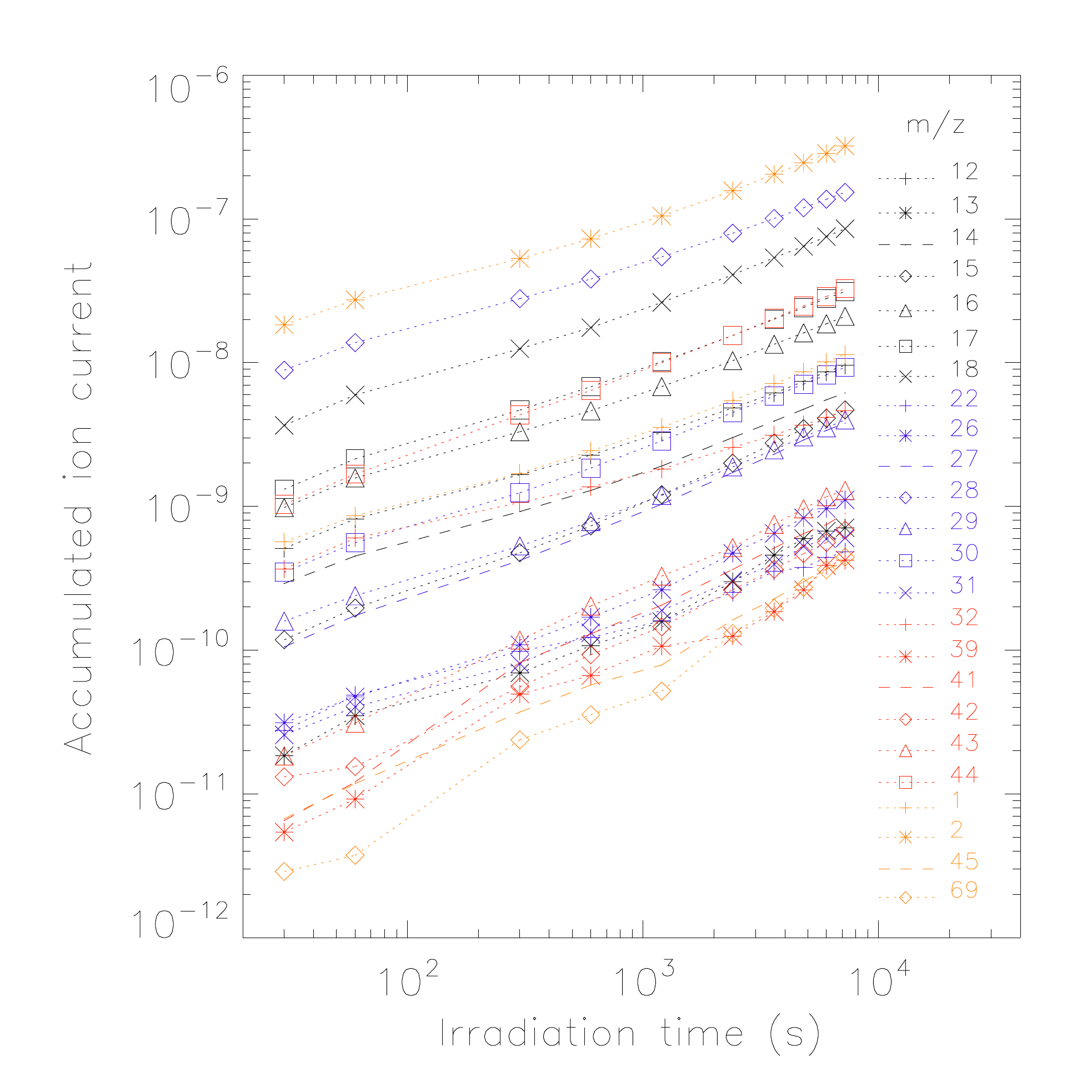} 
\caption{Accumulated QMS signal as a function of  the irradiation time for the photo-desorbing species detected during irradiation of BBs experiment (see Table~1.}
\label{fthree}
\end{figure}

\begin{deluxetable}{cc}
\label{tthree}
\tabletypesize{\scriptsize} 
\tablecaption{ Fragments detected by QMS during irradiation}
\tablewidth{0pt}
\tablehead{m/z & Fragments \\
 (amu) & }
\startdata
1  & H\\
2  & H$_2$ \\
12 & C\\
13 & CH\\
14 &  N, CH$_2$ \\
15 & HN, CH$_3$\\
16 & O, NH$_2$, CH$_4$\\
17 &  HO, NH$_3$\\
18 &  H$_2$O\\
22 &CO$_2^{++}$\\
26 & CN, C$_2$H$_2$\\
27 &HCN\\
28 & CO, N$_2$\\
29 & HCO, $^{13}$CO, $^{15}$NN, HN$_2$, CH$_3$N\\
30 & H$_2$CO, H$_2$N$_2$, CH$_4$N, C$_2$H$_6$, NO\\
31 & CH$_3$O,CH$_5$N, HNO\\
32 & O$_2$, CH$_3$OH, H$_2$NO, H$_4$N$_2$\\
39 & C$_2$HN,  C$_3$H$_3$ \\
41 & CH$_3$CN, C$_2$HO, CHN$_2$, C$_2$H$_3$N, C$_3$H$_5$\\
42 & NCO, N$_3$, C$_2$H$_2$O, CH$_2$N$_2$, C$_3$H$_6$\\
43 & HNCO, C$_2$H$_3$O, C$_2$H$_5$N, C$_3$H$_7$\\
44 & CO$_2$, N$_2$O, CH$_2$NO, C$_2$H$_4$O, CH$_4$N$_2$, C$_3$H$_8$\\
45 & $^{13}$CO$_2$, C$_2$H$_5$N, CH$_3$NO, CO$_2$H\\
69 & C$_3$H$_3$NO, C$_4$H$_5$O, C$_4$H$_7$N \\
\enddata
\end{deluxetable} 

In Table~3 are listed the molecular formula of the detected masses. Since the QMS is much more sensitive than FTIR spectrometer, we cannot exclude the contribution to each mass from species that are not clearly detected in the infrared spectra. In the following we discuss the major contributor for each mass. 

\begin{itemize}
\item $\rm m/z = 28$ ~\textendash~ This mass is directly associated with CO and N$_2$. The contribution of  CO$_2$ to mass 28 is less than 3\% since mass 44 is $\approx$ 5 times lower than m/z=28 and the relative intensity of $\rm m/z = 28$ in the CO$_2$ mass spectrum  is 12\%. The main fragment  of N$_2$ is atomic N ($\rm m/z = 14$) with a fragmentation ratio of 14\% with respect to N$_2$ (www.nist.gov); assuming that mass 14 comes exclusively from N$_2$, we conclude that the contribution of  N$_2$ to $\rm m/z = 28$ is at most 25\%. Thus, $\rm m/z = 28$ mostly comes from CO in the ice mixture.

\item $\rm m/z = 18$ ~\textendash~ This mass is assigned to H$_2$O which is the third most intense signal after m/z = 2 and 28 during irradiation. 

\item $\rm m/z=17$ ~\textendash~ OH is one of the carrier of this mass with an abundance of $\sim 26$\% with respect to water; however, this is not enough to justify the detected intensity. The ratio of m/z = 18 and 17 is $\sim  0.45$, a value much higher than that given for H$_2$O in Table~2. Another main contributor to this mass is NH$_3$, a parent molecule.

\item $\rm m/z = 44$ and 22  ~\textendash~ CO$_2$ is the most abundant product of the irradiated ice,  and the carrier of $\rm m/z = 44$. It also contributes to $\rm m/z = 12$, 16, 28, and 32. The mass 22 can be justified by a double ionization of CO$_2$ in the QMS with a ratio of 1.6\% respect to CO$_2$. N$_2$O possible carrier for the IR blended feature between 2200-2300 cm$^{-1}$ (see \citet{Pil10}) could also contribute to mass 44.

\item $\rm m/z= 43$ and 42  ~\textendash~ HNCO contributes to the infrared features in the $2300 - 2200$~cm$^{-1}$ range. Photo-desorption of this species can explain the signal of $\rm m/z = 43$ and 42 amu, HNCO and OCN respectively, \citep{Jim14}. Photo-desorption of these fragments have also been found during ultraviolet irradiation \citep{Che11}.

\item  $\rm m/z = 41$ and 39  ~\textendash~ Mass 41 has not been detected during ultraviolet irradiation of ice analogues \citep{Che11}. One possible candidate is CH$_3$CN (acetonitrile) which also contributes to mass fragment 39 and 40 amu. However, the intensity of the $\rm m/z = 40$ has not been observed (it is at the noise level). Acetonitrile has not been clearly detected in the infrared spectra. \citet{Men16} found that CH$_3$CN has a strong absorption at 2260~cm$^{-1}$ which overlaps with the HNCO and $^{13}$CO$_2$ features. QMS is more sensitive than FTIR allowing the detection of trace compounds. Alternatively, the $\rm m/z = 39$ can be  assigned to CHCN (cyanomethylene) which could also contribute to masses 13 and 26 amu. According to \citet{Mai88} CHCN has a strong feature at 1735~cm$^{-1}$ which could be blended with the  features around  the peak at 1688~cm$^{-1}$.

\item $\rm m/z = 32$ ~\textendash~ O$_2$ production can explain this mass.  Its fragmentation  also contributing to mass 16 (O). The photo-desorption of CH$_3$OH could also increase the $\rm m/z = 32$. However, direct photo-desorption of CH$_3$OH (from pure methanol ices) in the ultraviolet \citep{Dia16}, and X-rays (Chen, private communication) is negligible. On the other side, in ice mixture condition, CH$_3$OH could co-desorb with other molecules. However, the main signal of CH$_3$OH measured by QMS is $\rm m/z = 31$, whose intensity is significantly lower than $\rm m/z = 32$, while it should be the reverse (the relative intensity of mass 32 respect to mass 31 is 0.69). Thus, the methanol contribution to $\rm m/z = 32$ is very small.

\item  $\rm m/z = 31$, 30 and 29 ~\textendash~ direct CH$_2$OH and CH$_3$O fragments could contribute to mass 31. Direct photo-desorption of H$_2$CO (m/z = 30)  cannot explain the intensity of mass 29 (HCO). The contribution of CO  to mass 29 ($^{13}$CO) is 1.2\% smaller than the 10\% founds in our experiments. Thus, direct photo-desorption of HCO should take place. This species is also seen in the infrared spectra at 1848 cm$^{-1}$.  Nevertheless, more exotic compounds such as CH$_2$NH could be involved.

\item  $\rm m/z = 26$ and 27 ~\textendash~  Theses masses can be assigned to CN and HCN, respectively. Nevertheless, no HCN features have been detected in the infrared spectrum, probably due to the high reactivity of this species in the ice matrix. The detection of neutral reactive species suggest a non-dissociative photodesorption mechanism.

\item $\rm m/z = 14$ and 15  ~\textendash~  These masses are assigned to N and NH respectively. Mass 14 could come from N$_2$. However, mass 14, similar to mass 15, can also come from direct photo-desorption of NH$_3$ fragments formed in the ice.

\item $\rm m/z = 12$ and 16 ~\textendash~  CO, H$_2$O, NH$_3$ and CO$_2$ contribute to this signal. Carbon monoxide and dioxide participate to $\rm m/z=12$. However, a plethora of more complex products formed during irradiation could provide these fragments after ionization in the QMS.

\item m/z = 45. The CO$_2$ could contribute to mass 45 due to $^{13}$CO$_2$ which has a natural abundance of 1.1\% with respect to $^{12}$CO$_2$. From Figure~3 we can deduce a ratio $\rm (m/z=45)/(m/z=44) \approx 1.2 \%$, similar to the natural abundance. However, other compounds may contribute, such as NH$_2$CHO detected in the infrared spectrum (see Figure 1).

\item m/z = 69. Although noisy, this mass has been detected, suggesting photo-desorption of species with relatively high masses. Possible fragments are C$_3$H$_3$NO, C$_4$H$_5$O, or C$_4$H$_7$N. No signal has been found during the first minute of irradiation supporting a complex reaction processes in which precursors have to be formed.

\end{itemize}
A list of all the detected masses, and their possible assignments are reported in Table~3.

\begin{figure}
\centering
\includegraphics[width=18cm]{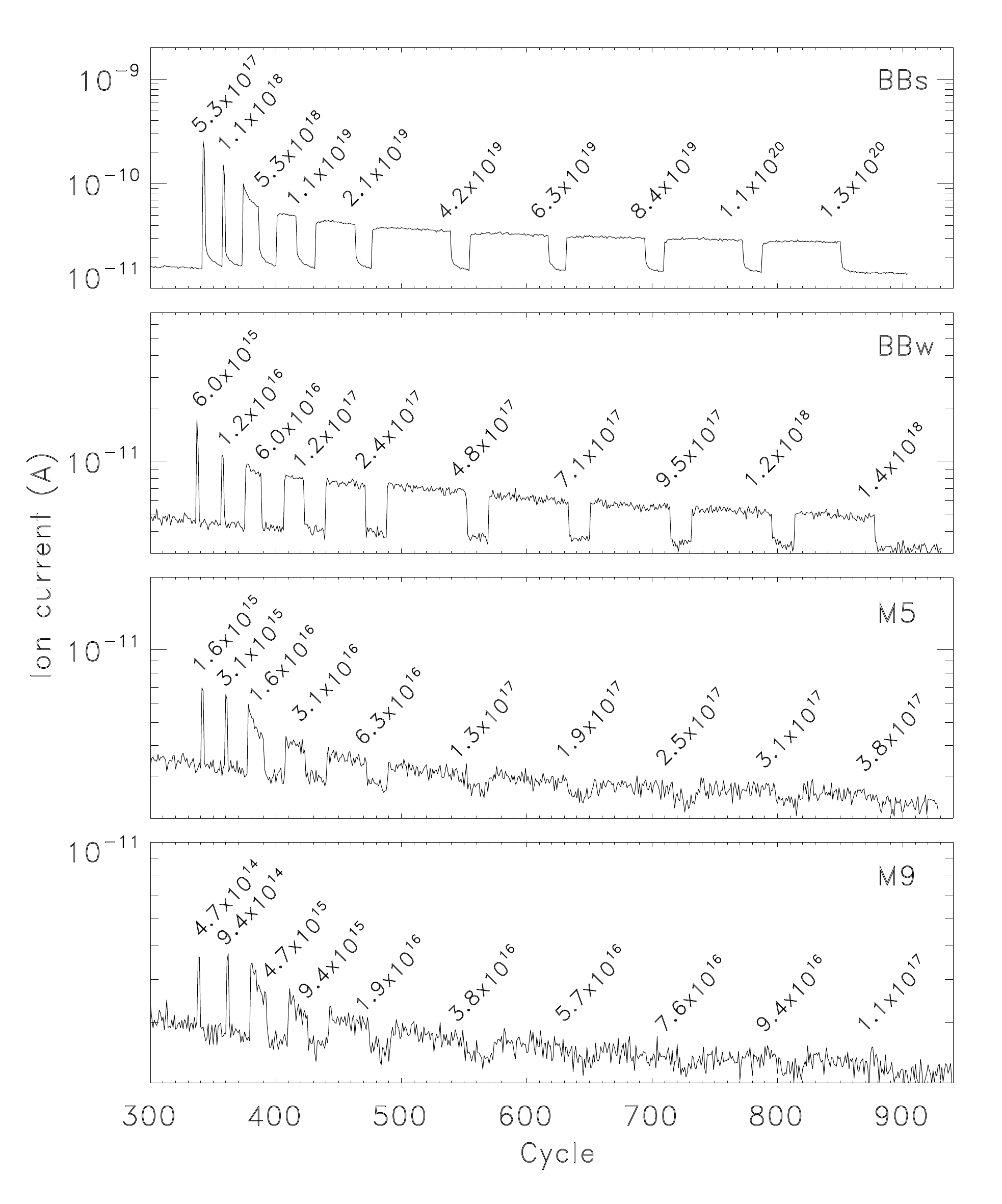} 
\caption{Ion current for $\rm m/z = 28$ detected by QMS during the experiments reported in Table~1. The impinging photon rate decreases from top to bottom. Labels indicate the total absorbed photons at the end of each irradiation step.}
\label{mass28}
\end{figure}

\section{Photo-desorption yield} \label{anal}
In the analysis of X-ray induced desorption, we use mass 28 as a reference, because this mass is the only detection common to all the experiments. The corresponding $\rm m/z = 28$ mass spectra are reported in Figure~\ref{mass28}. It is evident that the intensity of the ion current grows with the photon rate, and it decreases with the irradiation time. As a consequence in the experiments subjected to the lowest photon rates, the ion current intensity falls rapidly at the noise level. 

\begin{figure}
\centering
\includegraphics[width=18cm]{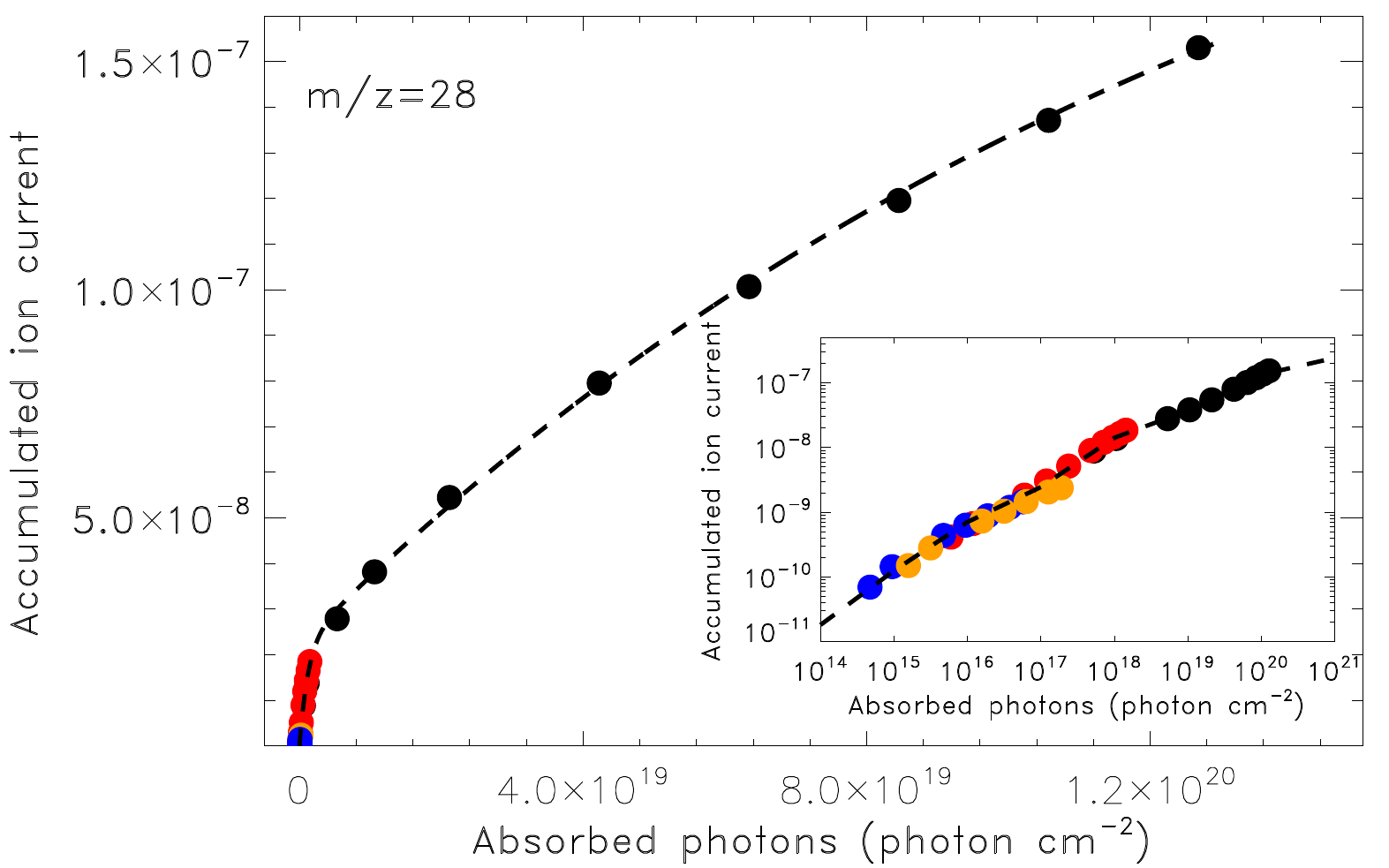} 
\caption{Accumulated QMS signal of m/z = 28 as a function of the absorbed photons during M5 (orange dots), M9 (blue dots), BBw (red dots), and BBs (black dots) experiments. The dashed line is the integral of the relation~(\ref{fit_2}). The inset panel shows the same data in logarithmic scale.}
\label{qms_str_tot}
\end{figure}

\begin{figure}
\centering
\includegraphics[width=18cm]{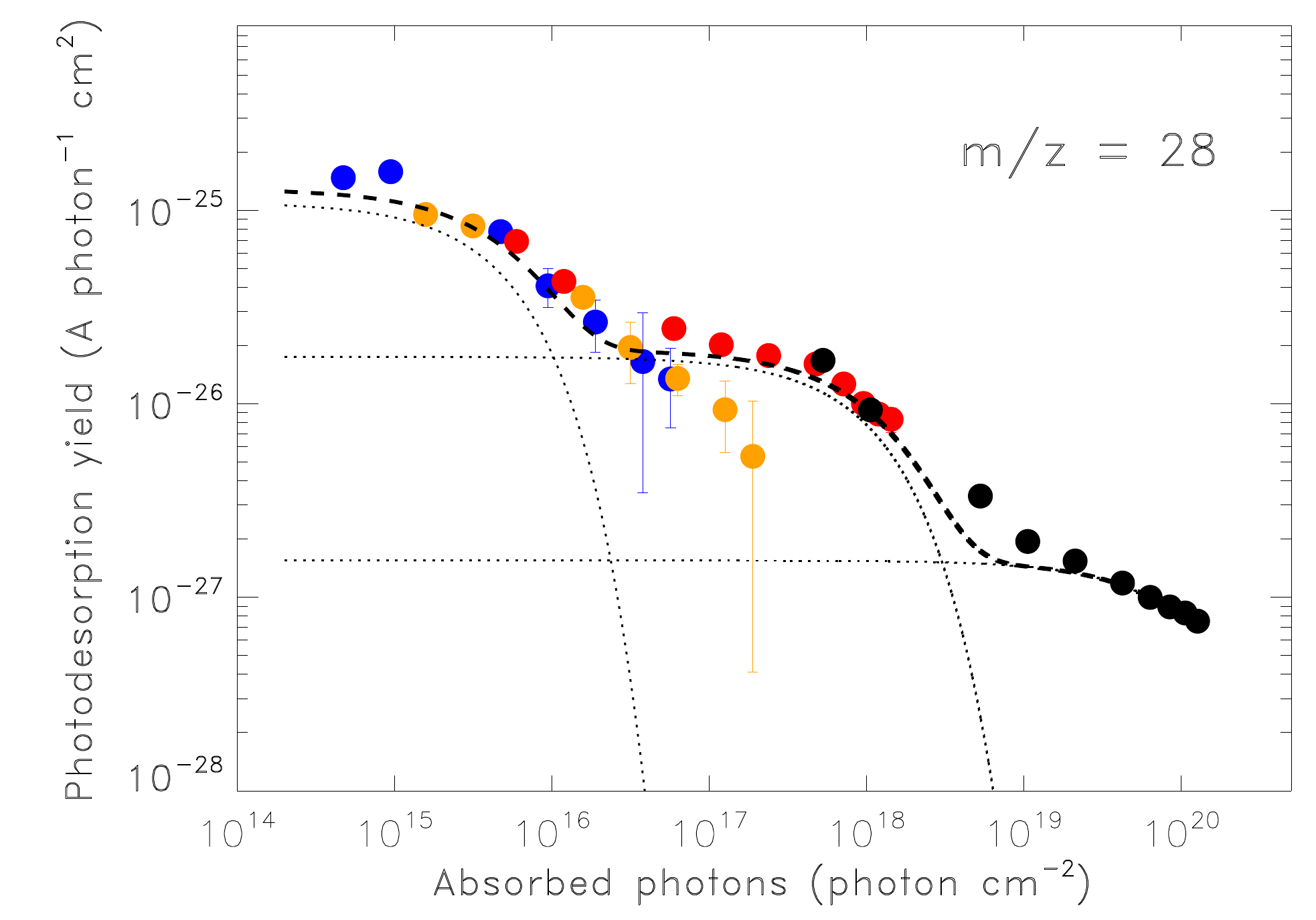} 
\caption{Photo-desorption yield vs the absorbed photons at each irradiation step for M5 (orange dots), M9 (blue dots), BBw (red dots),  and BBs (black dots)  experiments. Dashed line corresponds to equation~(\ref{fit_2}). Dotted lines represent single exponential contributions.}
\label{yield}
\end{figure}

To quantify the amount of species released in the photo-desorption events we show in Figure~\ref{qms_str_tot} the accumulated QMS ion current $IC$ of mass 28 (see Figure~\ref{mass28}) as a function of  the absorbed photons for all the experiments described in Table~1. We find that in a photon rate range covering three orders of magnitude, the accumulated ion current increases with the number of absorbed photons, $F_{\rm abs}$. This evidence is in contrast with the general trend of photo-reaction yields produced by X-ray irradiation in ices, in which  a strong dependence on the flux has been observed in many experiments involving e.g., CH$_3$OH \citep{Che13}, CO \citep{Cia06b}, and H$_2$O$-$CO mixtures \citep{Jim16}. Figure~\ref{qms_str_tot} also suggests that the photo-desorption yield is larger during the early stages of the experiments, decreasing with irradiation time, as it is actually confirmed by the data reported in Figure~\ref{yield}. We fit the photo-desorption yield of CO exploiting a superposition of three exponential functions 
\begin{equation}\label{fit_2}
Y_{\rm CO} = \frac{\Delta(IC)}
{\Delta F_{\rm abs}} = \sum_{i= 1}^3 A_i {\rm exp} \left( - \sigma_i {F}_{\rm abs} \right)
\end{equation}
where $\Delta (IC)$ is the ion current increment over each irradiation step, $\Delta F_{\rm abs}$ the corresponding number of absorbed photons per cm$^2$, $A_i$ and $\sigma_i$ are the pre-exponential factor and the desorption cross section, respectively. The major sources of error in the estimate of the accumulated ion current are noise and baseline. We perform the sum several times within limits varying randomly, producing a shift and a rotation of the baseline in either direction. The error on the number of absorbed photons has been chosen poissonian. Such errors have been propagated to obtain a $3 \sigma-$error on the yield, $Y_{\rm CO}$. Fitting parameters are reported in Table~4. This relation and its integral are reported as dashed lines in Figures~\ref{qms_str_tot} and \ref{yield}, respectively. 

Similarly to the ion current, the photo-desorption yield does not depend on the photon rate, although there might be a slightly departure from this trend at intermediate photon rates ($\sim {\rm few} \times 10^{12} - {\rm few} \times 10^{13}$~photon~s$^{-1}$), in the overlapping region ranging from $10^{16}$ to $10^{17}$ absorbed photons per square cm (see Figure~\ref{yield}). 

Since the curvature of the photo-desorption yield changes with the number of absorbed photon (and thus, for each photon rate with the irradiation time), several photo-desorption channels with different efficiencies might be coexistent, giving also rise to the modest dispersion at intermediate photon rates reported in Figure~\ref{yield}. To validate such hypothesis, we first investigate if the  decrease in the photo-desorption  yield is related to specific properties of the icy material such as thickness, composition, and  structure variations during the photo-processing.
\begin{deluxetable}{ccc}
\tabletypesize{\scriptsize} 
\tablecaption{Fitting parameters in equation~(\ref{fit_2})}
\tablewidth{0pt}
\tablehead{& $A$ & $\sigma$ \\
& (A $\times$ cm$^2$) & (cm$^2$)}
\startdata
1 & $1.0 \times 10^{-25}$ & $1.8 \times 10^{-16}$ \\
2 & $1.8 \times 10^{-26}$ & $8.1 \times 10^{-19}$ \\ 
3 & $1.6 \times 10^{-27}$ & $7.0 \times 10^{-21}$ 
\enddata
\label{tfour}
\end{deluxetable} 

\begin{figure}
\centering
\includegraphics[width=18cm]{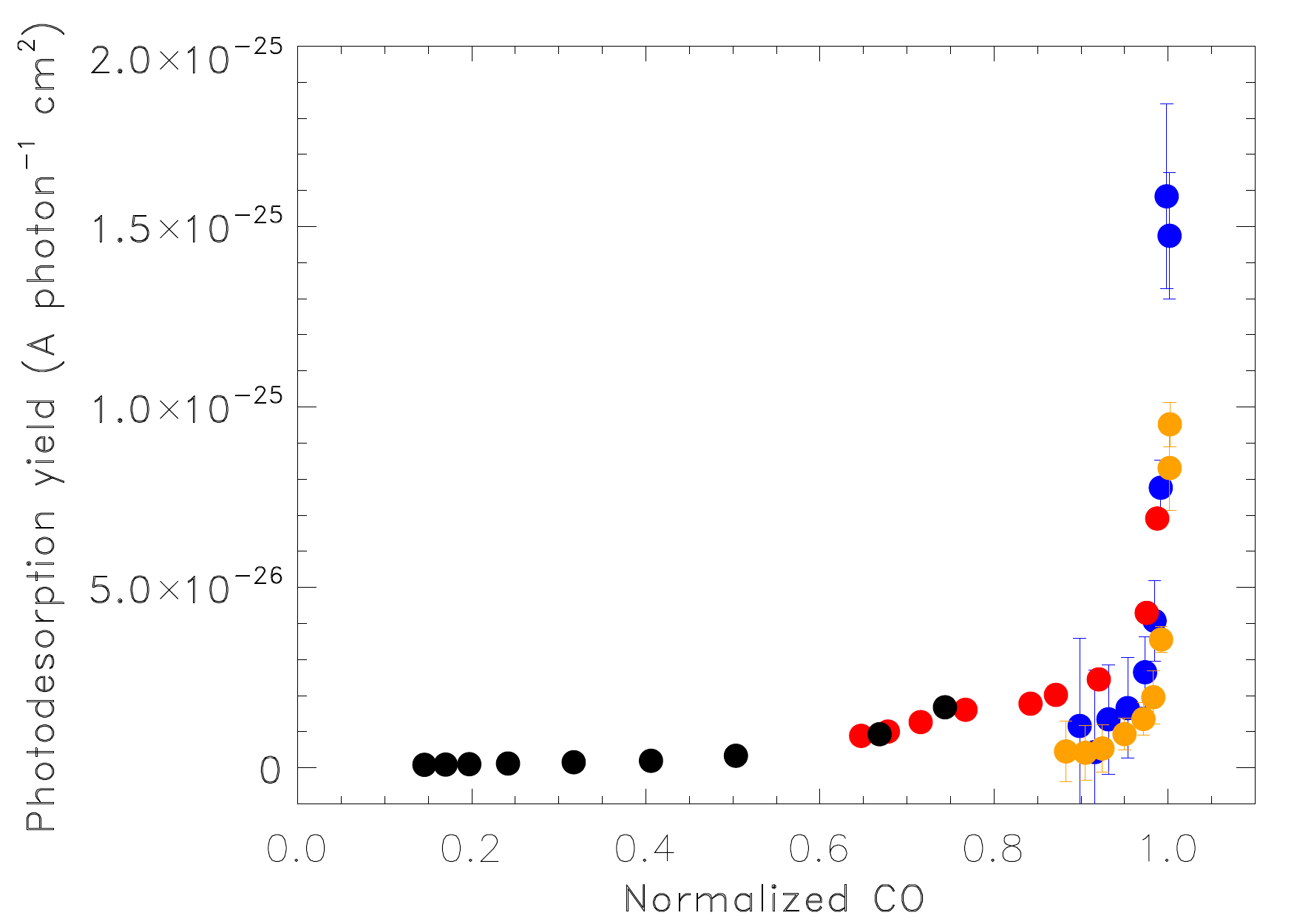} 
\caption{Photo-desorption yield of CO plotted against the CO column density in the ice. The column density is normalized to its value before processing. Different experiments are labeled as in Figure~\ref{qms_str_tot}.}
\label{fsix}
\end{figure}

We initially consider the case of thickness/composition, showing in Figure~\ref{fsix} the  CO photo-desorption yield plotted against the CO column density in the ice during the experiments. 

The use of CO is imposed by two different constraints on the other two parent molecules, namely: an uncertainty in the column density of H$_2$O where its infrared features overlap with that of NH$_3$, and QMS uncertainties for both NH$_3$ and H$_2$O, who share common fragments ($\rm m/z = 16$ and 17), thus preventing any accurate determination. It is interesting to note that while $Y_{\rm CO}$ decreases one order of magnitude, the ice CO column density falls at most 30\% depending on the experiment. For instance, during  M5 experiment, the CO column density decreases only 5\%. Considering CO$_2$ as main product, we may estimate that during the M5 experiment less than 3\% of the deposited CO is ejected into the gas-phase. A non-linear trend is common to all the experiments: this is in sharp contrast to the results of ultraviolet irradiation \citep{Ob07,Mu10}, in which the photo-desorption yield remains constant with the ice thickness up to the very last few monolayers. In addition, pure CO has a poor photochemistry, while the mixture in this paper leads to product formation, that affects the photo-desorption of CO. As X-ray processing continues, the mixture composition changes in time and therefore also the interactions between molecules; this was not the case in pure CO ices.

\begin{figure}
\centering
\includegraphics[width=18cm]{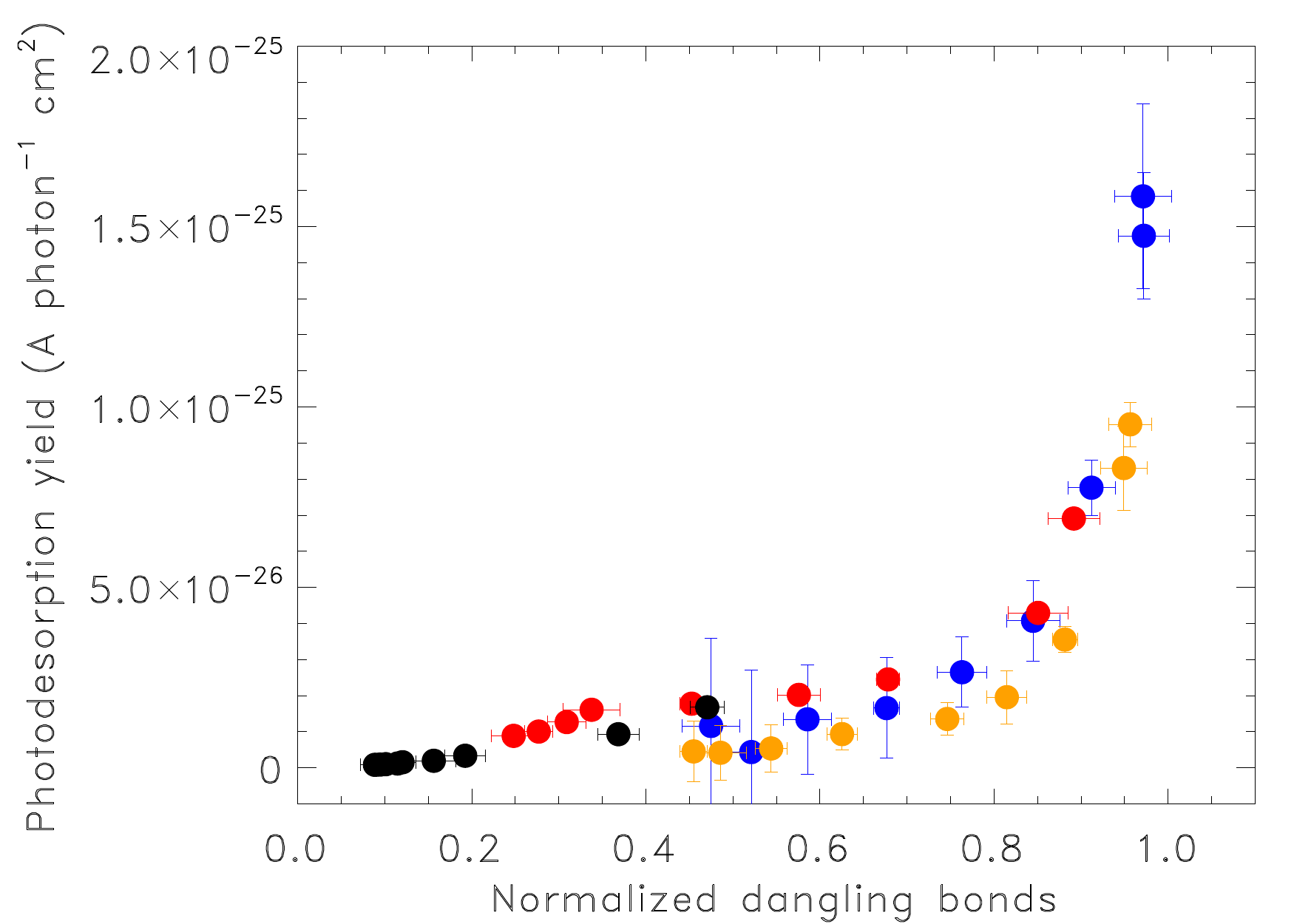} 
\caption{Photo-desorption yield of CO plotted against the $-$OH dangling bond infrared band at 3647~cm$^{-1}$. The band intensity is normalized to the its maximum intensity, occurring before the beginning of ice processing. Different experiments are labeled as in Figure~\ref{qms_str_tot}.}
\label{fseven}
\end{figure}

The  structure of water ices has been largely studied. The changes in the porosity of solid H$_2$O induced by thermal processing have been analyzed by \citet{Bos12}, who show the collapse of the porous structure by increasing the temperature. In addition, thermal desorption differences of $^{13}$CO and $^{15}$N$_2$ have been observed when deposited on compact or porous amorphous solid water \citep{Fay16}, requiring larger desorption temperatures in the amorphous case. In principle, the dangling modes are an indication of either pores in the ice bulk or a rough ice surface topology (e.g., \citealt{Pal06,Mei15}). However, the absence of dangling bond features in infrared spectra of astrophysical ice analogs, processed thermally or irradiated, does not imply a fully compacted film (\citealt{Mi17} and references therein). \citet{Rau07} studied the compaction of vapor-deposited amorphous solid water by energetic ions at 40 K. These authors found that the decrease in the strength of the $-$OH dangling features is related to a faster decrease of the surface area of porous ice with respect to the pore volume during the irradiation, i.e. the decrease of the pore internal surface area occurs before the collapse of the pore volume. We plot the $-$OH infrared feature at 3647~cm$^{-1}$ against the photo-desorption yield in Figure~\ref{fseven}.
We find a milder dependence of $Y_{\rm CO}$ on the $-$OH column density with respect to CO column density: for all the experiments, the photo-desorption yield decreases one order of magnitude when the strength of the dangling $-$OH features in the ice is reduced by at least 30\%. 

\section{Discussion}\label{ph_2}
The relation shown in Fig.~\ref{fsix} bears a resemblance to the results reported in \citet{Ber06}, who performed pulsed laser photo-desorption experiments of a few monolayer of amorphous water ice deposited over graphite and platinum. These authors found a high increase of the photo-desorbed molecules followed by a fast decay, that disappeared after a few seconds of irradiation. In addition, they observed a limited decrease of the ice coverage. \citet{Ber06} speculated that the observed phenomenon might results from multiphoton excitation leading to formation of excitons in the ice, that eventually promote desorption of neutral and ionic species from adsorbed layers of H$_2$O. 

The absorption of X-rays in the ice produces the ejection of energetic electrons, which in turn gives rise to a secondary electron cascade. The interactions of secondary electrons result in ionizations and excitations, i.e. excitons, of the molecules in the solid. In water ice, excitons have energies in the 8 to 14 eV range, and they can move up to tens of monolayers from the location at which they are generated \citep{MC18}. The exciton energy of CO is in the same range as water (the ionization energy is around 14 eV). \citet{MC18} show that the electron desorption intensity decreases with increasing thickness of a C$_6$H$_6$ layer deposited over a water ice, suggesting that water is the major agent in the process. On the other side, ultraviolet photo-desorption of CO ice is quenched by about one order of magnitude when CO molecules interact with a water ice surface \citep{B12}. In our experiment the interactions between radiation and ice molecules are mediated by secondary electrons generated in primary photo-absorptions events, whose effects are rather similar in the interaction with CO and H$_2$O. Thus, both species should participate in the desorption of weakly bound species. This reflects the very different nature of the interactions of ultraviolet radiation and X-rays with the ice.

In our experiment the whole picture depends on the the X-ray transmission within the ice, with absorption events ~\textendash~ that depend on the photon energy ~\textendash~ occurring mainly in the bulk of the sample. Excitons generated within the so-called selvedge region are a source of desorption through DIET, in which e.g., a hot water molecule may escape from the ice after having inelastically scattered up to the surface. H$_2$O may also fragment giving rise to H and OH, also likely to be hot. Such species may escape directly into the gas-phase, or combine with other mobile hydrogen atom producing H$_2$ and H$_2$O, which in turn may escape. Such scattering events may also efficiently induce the desorption of weakly bound adsorbates at the surface, e.g. H$_2$O, and CO molecules. In the bulk, exciton relaxation induces the formation of new species through electronic excitation pathways. 

The salient points of our results are as follows: (1) the ion current depends on the flux of incoming photons, independently on the photon energies (in the range exploited in this work); (2) the accumulated (over single irradiation shift) ion current $IC$ increases with the number of absorbed photons; (3) the $IC$ profile shows (at least) two regimes, an initial steep rise followed by a slower increase with the number of absorbed photons; (4) the photo-desorption yield $Y$ decreases with the number of absorbed photons, and with the photon rate; (5) $Y$ changes its curvature few times in the absorbed photon range used in the experiments; (6) $Y$ initially decreases about one order of magnitude in the face of a very moderate decrease of the CO column density, and then decreases much more slowly; (7) $Y$ shows a similar trend when plotted against the $-$OH infrared feature at 3647~cm$^{-1}$ in the ice, although the resulting profile is smoother than in the previous case.    

Points (3) and (5) may reflect the presence of different coexistent photo-desorption mechanisms,  and/or variations in the structural properties of the ice. The X-ray production of excitons in the selvedge provides an efficient desorption mechanism for weakly bound adsorbates at the solid ice surface (substrate-mediated exciton-promoted desorption, see \citealt{MC18}). Laboratory measurements of electron low-energy irradiation of benzene (weakly) bound to the surface of an amorphous water ice film give an effective desorption cross-section $\sim 10^{-15}$~cm$^2$ \citep{M16}. Such value is consistent with our description of the CO photodesorption yield at low number of absorbed photons $\sim 1.8 \times 10^{-16}$~cm$^2$ (see Table~4). If ${\cal N}_{\rm wb}$ is the number of weakly bound molecules on the ice surface, the supply of excitons, ${\cal N}^\star_{\rm exc}$ provided by X-ray absorption induces a steady desorption of weakly bound molecules until ${\cal N}^\star_{\rm exc} \sim {\cal N}_{\rm wb}$. At that stage, the substrate-mediated exciton-promoted desorption yield falls rapidly leaving a residual tail of species desorbing from the substrate. This desorbing rate is due to both excitons and direct PSD, and/or other kind of photon-induced fragmentation events. This scenario is supported by the decrease of the photo-desorption yield with the photon rate: if ${\cal N}$ is the number of molecules close to the substrate surface, then 
\begin{equation}
Y \sim \frac{\Delta \left( {\cal N}_{\rm wb} + {\cal N} \right)}{\Delta F_{\rm abs}} \to \frac{\Delta {\cal N}}{\Delta F_{\rm abs}}
\end{equation} 
as soon as ${\cal N}_{\rm wb}/\Delta F_{\rm abs} \to 0$.  At this stage the photodesorption cross-section is decreased by about a factor 20 (see Table~4, second exponential function), consistently with the cross-section for the loss of molecular water from solid water, $\sim 10^{-17}$~cm$^2$ derived by \citet{M16}. This transition occurs when $F_{\rm abs} \sim 2 \times 10^{17}$~photons~cm$^{-2}$, when a plateau in $Y$ occurs until porosity effects, decreasing the surface/volume ratio, produce the ultimate decline of the yield shown in Fig.~\ref{fseven}. This final transition happens at $F_{\rm abs} \sim 2 \times  10^{18}$~photons~cm$^{-2}$, corresponding to $Y \sim 1 \times 10^{-26}$~A~photon$^{-1}$~cm$^2$ (see Fig.~\ref{fseven}).

Finally, we consider the possibility of the contribution of local heating to desorption; when local heating contributes to desorption, the photodesorption yield cannot be a linear relation to photon flux. Unfortunately, we performed only two different photon flux irradiations with the same ice thickness. However, preliminary results with a pure CO ice, shows that the within an order of magnitude in photon flux ($\sim 10^{13} - 10^{14}$~photons~s$^{-1}$) the photodesorption yield is directly proportional to photon flux, excluding therefore a relevant role of local heating in photodesorbing the ice sample.

\section{Conclusions}
In this work we study the photo-desorption arising from soft X-ray irradiation of an ice mixture containing water, carbon monoxide, and ammonia (1.5:1.3:1 in ratio). Two distinct regimes apply: an initial phase at low photon fluence in which substrate-mediated exciton-promoted desorption dominates the photo-desorption yield inducing the release of weakly bound (to the surface) species, followed by the release of species directly from the substrate via exciton-promoted and fragmentation channels, at much higher fluxes. The transition between these two regimes is regulated by the number of weakly bound species.  

The present experiment bears implications for X-ray photo-desorption in space. On the base of data provided by the present experiments, we may derive an upper limit to the CO photo-desorption yield. Assuming that the difference between photolized CO and new products $\Delta N_{\rm C}$ is due to photo-desorption, we may convert the photo-desorption yield measured in ($A/ \rm photon)~cm^{2}$ into a photo-desorption yield in $\rm molecule~photon^{-1}$. In the first five minutes of irradiation (thus regarding just the experiments M3, M5, and BBw), CO$_2$ and HCO are the only C-bearing products observed in the infrared. If we define $AC^\star$ the accumulated ion current in these initial five minutes of irradiation we obtain the yield in molecules photon$^{-1}$:
\begin{equation}\label{fit_3}
Y^{\rm m}_{\rm CO} = \left( \frac{\Delta N_{\rm C}}{AC^\star} \right) Y_{\rm CO} = 9.5 \times 10^{24} \, Y_{\rm CO}
\end{equation}
where $Y_{\rm CO}$ the photo-desorption yield given in equation~(\ref{fit_2}). The photo-desorption yield in our experiments varies from 1.4 to 0.007 molecule photon$^{-1}$ in the range $\sim 10^{15} - 10^{20}$~absorbed photons cm$^{-2}$. For comparison, the photo-desorption yield of CO molecules in a CO ice at 15 K irradiated by ultraviolet is $\sim$ 3.5 $\times$ 10$^{-2}$ molecules photon$^{-1 }$ \citep{Mu10}.
\begin{figure}
\centering
\includegraphics[width=18cm]{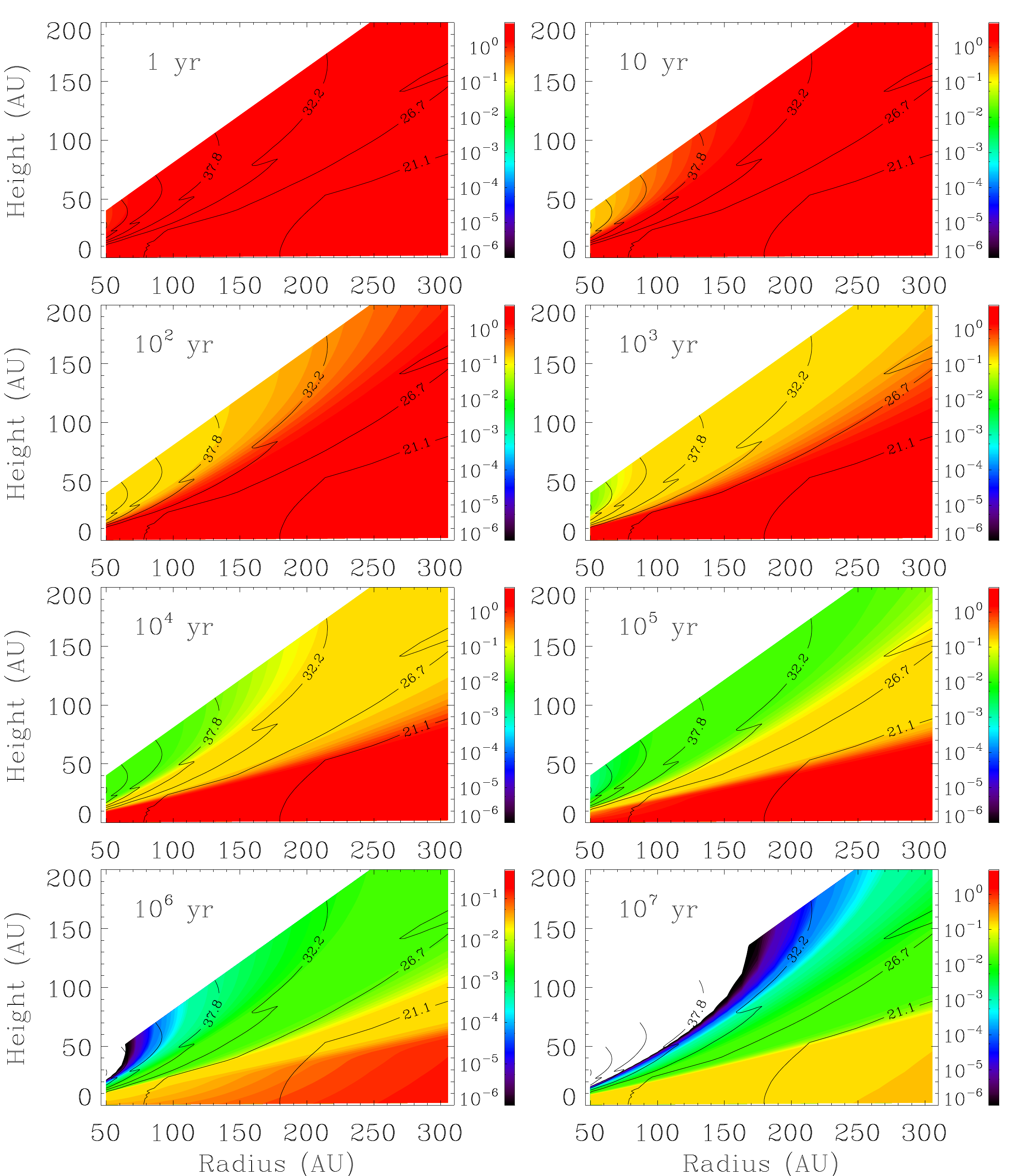} 
\caption{Photo-desorption yield of CO in molecule~photon$^{-1}$ as function of the evolutionary ages of a synthetic protoplanetary disk around a classical T Tauri star \citep{W12}. The photo-desorption yield of CO has been computed assuming the same yield as in the X-ray experiment, see equation~(\ref{fit_3}).}
\label{proto_d}
\end{figure}

Using our results we can simulate the evolution of the photo-desorption yield in a protoplanetary disk, under the assumptions that photo-desorption does not depend on the ice thickness, and that accretion onto the grains is inhibited during the simulation. Inserting equation~(\ref{fit_3}) in the protoplanetary disk model by \citet{W12} we simulate CO X-ray induced photo-desorption yield in a T~Tauri irradiated disk (see Fig.~\ref{proto_d}). The photo-desorption yield results $\sim 1.2$ molecules photon$^{-1}$ in the mid-plane of the disk during the first million years. Interestingly, since the short-lived exciton-promoted desorption phase depends on the intensity of the impinging radiation, after a short transient the photo-desorption yield becomes much weaker in the outer regions of the disk than in the mid-plane.

X-ray may induce sublimation of ice, and therefore desorb molecules from grains, through the so-called spot-heating, in which desorption is due to transient heating events (e.g., \citealt{N01}). In fact, we recorded an enhancing of temperature at the window during the X-ray irradiation. Although, the increase in temperature is confined within 1~K, the effective temperature increase of the system ice/grain may be much higher, as the window is tightly connect with the cryostat. We shall discuss this issue in a future work.

We conclude that X-ray photodesorption processes should be taking into account in the modelling of protoplanetary disks, and more in general in X-ray dominated regions. 

\section*{Acknowledgements}
We acknowledge the NSRRC general staff for running the synchrotron radiation facility. We also thank Dr. T.-W. Pi, the spokesperson of BL08B in NSRRC.

This work has been supported by the project PRIN-INAF 2016 The Cradle of Life - GENESIS-SKA (General Conditions in Early Planetary Systems for the rise of life with SKA). We also acknowledge support from INAF through the “Progetto Premiale: A Way to Other Worlds” of the Italian Ministry of Education, the MOST grants MOST 107-2112-M-008-016-MY3 (Y-JC), Taiwan, and project AYA-2011-29375, AYA- 2014-60585-P of Spanish MINECO.

\end{document}